\documentclass[onecolumn,showpacs,preprintnumbers,amsmath,amssymb]{revtex4}

\def\be{\begin{equation}}
\def\ee{\end{equation}}
\def\ba{\begin{eqnarray}}
\def\ea{\end{eqnarray}}
\def \bea{\begin{eqnarray}}
\def \eea{\end{eqnarray}}

\usepackage{graphics}
\usepackage{graphicx}
\usepackage{dcolumn}
\usepackage{bm}
\usepackage{epsfig}
\usepackage{graphicx}
\usepackage{multirow}
\usepackage{color}
\usepackage{dcolumn}
\usepackage{graphicx,epsfig}%
\usepackage{subfig}
\def \ee{\end{equation}}
\def \be{\begin{equation}}
\def \bea{\begin{eqnarray}}
\def \eea{\end{eqnarray}}

\preprint{}
\begin{document}

\title{A scale dependent black hole in three-dimensional space-time}

\author{Benjamin Koch, Ignacio A. Reyes, and \'Angel Rinc\'on}
 \affiliation{
Instituto de F\'{i}sica, \\
Pontificia Universidad Cat\'{o}lica de Chile, \\
Av. Vicu\~{n}a Mackenna 4860, \\
Santiago, Chile \\}
\date{\today}

\begin{abstract}
Scale dependence at the level of the effective action is a generic result of quantum field theory.
Allowing for scale dependence of the gravitational couplings
leads to a generalization of the corresponding field equations.
In this work, those equations are solved by imposing the ``null energy condition'' in three-dimensional space time
with stationary spherical symmetry. The constants of integration are
given in terms of the classical BTZ parameters plus one additional
constant, that parametrizes the strength of the scale dependence.
The properties such as asymptotics, horizon structure, and thermodynamics are discussed.
It is found that the black hole entropy shows a remarkable transition from the
usual ``area~law'' to an ``area~$\times$~radius'' law.
\end{abstract}

\maketitle

%
\section{Introduction}

Gravity in $(2+1)$ dimensions is a vibrant field of research. 
This is in part due to the fact that the absence of propagating degrees of freedom
makes things simpler than in $(3+1)$ dimensions, in particular when dealing with the challenge of formulating
a quantization of this theory. Another
important feature of gravity in $(2+1)$ dimensions is the deep connection to Chern-Simons
theory \cite{Witten:1988hc,Achucarro:1987vz,Witten:2007kt}.
This by itself makes the black hole solution \cite{Banados:1992wn,Banados:1992gq} found by Ba\~nados, Teitelboim, and Zanelli (BTZ)
an extremely interesting research object, which has been generalised in many directions. 
An additional component that motivates the research on black holes in three dimensions
is their prominent role in the context of the AdS/CFT correspondence~\cite{Maldacena:1997re,Strominger:1997eq,Balasubramanian:1999re,Aharony:1999ti}.

Despite of some progress, the consistent formulation of quantum gravity remains an 
open task  which is attacked in many different ways \cite{Wheeler:1957mu,Deser:1976eh,Rovelli:1997yv,Bombelli:1987aa,Ashtekar:2004vs,Sakharov:1967pk,Jacobson:1995ab,Verlinde:2010hp,Reuter:1996cp,Litim:2003vp,Horava:2009uw,Charmousis:2009tc,Ashtekar:1981sf,Penrose:1986ca,Connes:1996gi,Nicolini:2008aj,Gambini:2004vz} (for a review see \cite{Kiefer:2005uk}).
Even though many approaches to quantum gravity are very different, most of them have the common feature that
the resulting effective action of gravity acquires a scale dependence.
This means that the couplings appearing in the quantum- effective action
(such as Newtons coupling $G_0$, or the cosmological term $\Lambda_0$)
become scale dependent quantities ($G_0 \rightarrow G_k$, $\Lambda_0\rightarrow \Lambda_k$).
There is quite some evidence that this scaling behavior is in agreement with
Weinberg's Asymptotic Safety 
program~\cite{Weinberg:1979,Wetterich:1992yh,Dou:1997fg,Souma:1999at,Reuter:2001ag,Fischer:2006fz,Percacci:2007sz,Litim:2008tt}.
In particular, the effective action and running couplings in three dimensions have been studied in \cite{Avramidi:2000bm,Demmel:2012ub}.
In any case, 
scale dependent couplings can be expected to
produce differences to classical general relativity, such as modifications
of classical black hole backgrounds~\cite{Nojiri:1996vs,Bonanno:1998ye,Bonanno:2000ep,Bonanno:2006eu,Reuter:2006rg,Koch:2007yt,Burschil:2009va,Falls:2010he,Cai:2010zh,Litim:2011cp,Nicolini:2011nz,Becker:2012js,Becker:2012jx,Falls:2012nd,Contreras:2013hua,Koch:2013owa,Koch:2013rwa,Koch:2014cqa,Koch:2015nva,Rodrigues:2015rya,Gonzalez:2015upa}.

In this paper the possible effects of scale dependence on the black hole in three dimensional gravity will be
investigated in the light of the effective action approach.
We will use the scale-field method applied to the Einstein-Hilbert truncation,
which allows to derive generalized Einstein equations for the case
of scale dependent couplings \cite{Reuter:2003ca,Koch:2010nn,Domazet:2012tw,Koch:2014joa}.
The theoretical uncertainty concerning the functional form of the scale dependence
of $G_k$ and $\Lambda_k$ will be avoided. Instead, the most general stationary
spherically symmetric solution without angular momentum, which 
is in agreement with the common ``null energy condition''
for the effective stress energy tensor, will be derived. 
It is further shown that this solution corresponds also to the most general case
which is in agreement with the ``Schwarzschild relation'' $g_{tt}=-1/g_{rr}$.

The paper is organized as follows:
In subsection \ref{sec_RNclas} a small collection of basic properties
of the classical BTZ solution is presented. In subsection \ref{sec_scaleset}
the concept of effective action with scale dependent couplings is reviewed.
In section \ref{sec_BTZsol} those techniques will be used to derive and discuss
a new black hole solution in three dimensions with scale dependent couplings. 
In subsection \ref{sec_NullEn} the ``null energy condition'' for the effective stress
energy tensor is formulated and its connection to the ``Schwarzschild relation'' is reviewed.
The solution is presented in subsection \ref{sec_findBTZ}, 
in subsection \ref{sec_BTZasympt} the asymptotic behavior of the solution is discussed,
the horizon structure is analyzed in subsection \ref{sec_BTZhor}, and
the thermodynamic properties are discussed in subsection \ref{BTZthermo}.
Results are summarized in section \ref{sec_sum}.

\subsection{The classical BTZ solution without scale dependence}
\label{sec_RNclas}

In this subsection the key features such as line element and thermodynamics 
of the classical BTZ black hole solution \cite{Banados:1992wn,Banados:1992gq}
will be listed. This summary will be limited to the case of zero angular momentum.

In the metric formulation the gravitational action in three dimensions
\be\label{actionBTZ}
{\mathcal{S}}(g_{\mu \nu})=\int d^3x \sqrt{-g}\frac{\left( R-2\Lambda_0 \right)}{G_0},
\ee
gives the equations of motion
\be\label{eomBTZ}
 G_{\mu\nu}=-g_{\mu\nu}\Lambda_{0},
\ee
where $\Lambda_0$ is the cosmological constant
and $G_0$ is Newton's constant.
For the non rotating BTZ solution, 
the line element takes the form 
\be\label{lineele}
ds^2= -f_0(r) \, dt^2+ f_0(r)^{-1} \, dr^2 + r^2 d\phi^2,
\ee
with
\be
f_0(r)=-G_0 M_0 + \frac{r^2}{\ell_0^2},
\ee
where $\Lambda_0=-1/\ell_0^2$ and $M_0$ is the mass of the black hole.
For this solution the black hole entropy and temperature read
\be
S_{BTZ}= 4 \pi \ell_0\sqrt{\frac{M_0}{G_0}}\ ,\hspace{2cm}
T_{BTZ}=\frac{\sqrt{M_0G_0}}{2 \pi \ell_0} .
\ee

%
\subsection{Scale dependent couplings}
\label{sec_scaleset}

This subsection summarizes the equations of motion for the scale dependent
space-times in three dimensions.
The notation and scale setting procedure is according to \cite{Reuter:2003ca,Koch:2010nn,Domazet:2012tw,Koch:2014joa,Contreras:2016mdt}.

The scale dependent effective action is
\be\label{actionEG}
\Gamma(g_{\mu \nu}, k)=\int d^3x \sqrt{-g}\frac{\left( R-2\Lambda_k \right)}{G_k}.
\ee
By varying~(\ref{actionEG}) with respect to the metric field one obtains

\begin{eqnarray}
 G_{\mu\nu}=-g_{\mu\nu}\Lambda_{k}
 +8 \pi G_k T^{}_{\mu \nu}.
\label{eomg}
\end{eqnarray}
The effective stress energy tensor $T^{}_{\mu \nu}$
contains the actual matter contribution $T^{m}_{\mu \nu}$ and a contribution $\Delta t_{\mu \nu}$
induced by the possible coordinate dependence of $G_k$~\cite{Reuter:2003ca}
\be\label{Teff}
T^{}_{\mu \nu}=T^{m}_{\mu \nu}-\frac{1}{8 \pi  G_k}\Delta t_{\mu \nu},
\ee
where
\be
\Delta t_{\mu \nu}=G_{
k}\left(g_{\mu\nu}\Box-\nabla_\mu\nabla_\nu\right)\frac{1}{G_{k}}. \label{dt}
\ee
By varying~(\ref{actionEG}) with respect to the scale-field $k(x)$
one obtains
the algebraic equations
\be\label{eomk}
\left[R \frac{\partial}{\partial k} \left(\frac{1}{G_k}\right)-
2 \frac{\partial}{\partial k}\left(\frac{\Lambda_k}{G_k}\right)\right]
=0.
\ee
The above equations of motion are consistently complemented
by the Bianchi identity, reflecting invariance under 
coordinate transformations
\be\label{diffeo}
\nabla^\mu G_{\mu \nu}=0.
\ee

\section{Scale dependent solution without angular momentum}
\label{sec_BTZsol}

Let us now turn to solving the system of equations~(\ref{eomg}-\ref{diffeo})
assuming a stationary space-time with rotational symmetry and no angular momentum.
The most general line element in agreement with this symmetry is
\be\label{lineelegen}
ds^2= -f(r) \, dt^2+ g(r) \, dr^2 + r^2 d\phi^2.
\ee
Apart from the two functions $f(r)$ and $g(r)$, the system has to be solved
for the scale field $k(r)$.
In principle this is possible,
as soon as the functional form of the scale
dependent couplings $G_k$ and $\Lambda_k$ is known, 
for example from some Functional Renormalisation Group (FRG) equation.
Those functions have been calculated by using 
various methods and approximations. 
However, it has up to now not been possible
to obtain an exact and scheme independent expression of the effective average action.
Therefore, the functional form of $\Lambda_k$ and $G_k$
is subject to very large theoretical uncertainties.
This problem is aggravated by the fact
that most functional forms of $\Lambda_k$ and $G_k$
are either only valid in the UV or in the IR.
Given those drawbacks we will proceed with a method
that has been previously applied in four dimensions \cite{Reuter:2003ca,Koch:2010nn,Domazet:2012tw,Koch:2014joa}:
The first step is to realize that the only appearance of
the scale field $k(r)$ is within the couplings and that
for any solution of the
system the functions $\Lambda_k$ and $G_k$
will inherit a radial dependence from $k(r)$.
Thus, one might try to solve the system
for $\left\{f(r),\, g(r), \, \Lambda(r),\, G(r)\right\}$ 
(instead of solving for $\left\{f(r),\, g(r), \, k(r)\right\}$).
However, since one dealt one unknown function $k(r)$
for two unknown functions $\Lambda(r)$ and $G(r)$,
the system is underdetermined.
In order to obtain a determined system again   
one has to impose an additional condition. 

We therefore stress that in this approach, we shall make no attempt to fix the renormalization scale $k=k(r)$. Our strategy is converse: regardless of the specific form of $k(r)$, any running-coupling solution will inherit a spatial coupling dependence. Any solution to Einstein's equations that is static, spherically symmetric and fulfills the null-energy condition (see below) must belong to the family of configurations described below. They are determined by four integration constants.

\subsection{The null energy condition (NEC)}
\label{sec_NullEn}

The most common type of conditions in classical general relativity 
are energy conditions~\cite{Wald:1984rg,Rubakov:2014jja,Parikh:2015ret}, where one typically 
distinguishes between  the dominant, weak, strong, and null condition.
The less restrictive of those conditions is the null condition, which states
that for a null vector field $l^\nu$ the matter stress energy tensor satisfies
\be\label{nullcond}
T^{m}_{\mu \nu}l^\mu l^\nu\ge0.
\ee 
Since we are interested in black hole solutions it is crucial to note
that the energy condition (\ref{nullcond}) is actually necessary
for the proof of fundamental black hole theorems
such as the no hair theorem~\cite{Heusler:1996ft} or the second law of black hole
thermodynamics~\cite{Bardeen:1973gs}.
Therefore, if one is looking for black hole solutions that
are in agreement with those two fundamental theorems, it is natural to impose
that appearance of scale dependence does not spoil or alter this property
for the effective stress energy tensor 
\be\label{nullcond1}
T_{\mu \nu}l^\mu l^\nu \overset{!}{=} T^{m}_{\mu \nu}l^\mu l^\nu \ge0.
\ee 
A physical interpretation of this condition is that one imposes
that not even a light-like observer can observe a difference between
the energy density due to the presence of matter and the effective energy density
due to the combined matter and scale dependence effects.
The relation (\ref{nullcond1}) holds if one maintains the standard matter condition (\ref{nullcond}) and one further imposes that the extra contribution to the stress tensor \eqref{dt} induced by the variation of the couplings satisfies
\be\label{nullcond2}
\Delta t_{\mu \nu}l^\mu l^\nu=0.
\ee
In a spherically symmetric setting one can solve this condition for the scale dependent coupling (\ref{lineelegen})
without the use of the equations of motion (\ref{eomg}) giving
\begin{align}
G(r) &= a\left[ \int_{r_0}^{r}\sqrt{f(r') \cdot g(r')} \ dr' \right]^{-1},
\end{align}
where $a$ and $r_0$ are constants.
The next step consists in finding the metric functions $f(r)$ and $g(r)$
which appear in this integral.
This can be achieved by a straight forward argument following Jacobson~\cite{Jacobson:2007tj}:
one can choose the null vector field to be
$l^\mu=\left\{\sqrt{g},\sqrt{f},0\right\}$.
Combining the equations of motion (\ref{eomg}) for this vector field
with the condition (\ref{nullcond1}) gives
in regions without external matter $(T^m_{\mu \nu}=0)$ 
\be
R_{\mu \nu}l^\mu l^\nu=(f \cdot g)' \frac{1}{2 r g}=0
\ee
and thus $f \sim 1/g$. By making 
use of time reparametrization invariance, this allows to write $f(r)=1/g(r)$,
which corresponds to the so called Schwarzschild relation.
It is interesting to note that this 
common relation further ensures that the radius coordinate is an affine parameter on the radial null geodesics \cite{Jacobson:2007tj}.
With this relation the line element is
\be\label{lineelans}
ds^2= -f(r) \, dt^2+ f(r)^{-1} \, dr^2 + r^2 d\phi^2
\ee
and the equations of motion can be completely solved for the three functions
$\left\{f(r), \, \Lambda(r), \,G(r)\right\}$.

The necessity of imposing an additional condition arises due to the fact that 
we avoid using an ansatz for $G_k$ and $\Lambda_k$, (with an additional field variable $k(r)$). Instead we are
dealing directly with $G(r)$ and $\Lambda(r)$ and thus need an additional constraint. 
In \cite{Contreras:2013hua}, for the case of a spherically symmetric solutions, this additional constraint was chosen to be the
usual Schwarzschild relation $f\cdot g \equiv 1$, 
which can be derived from the null energy condition (\ref{nullcond2}).
However, as discussed above, the null energy condition has a stronger physical motivation and a broader applicability, allowing to go 
beyond spherically symmetric black holes.

\subsection{A non-trivial solution for scale dependent couplings}
\label{sec_findBTZ}

Based on~(\ref{nullcond1}) one finds that the equations~(\ref{eomg}) are solved by
\bea
G(r)&=&\frac{G_0^2}{G_0+\epsilon r(1+G_0 M_0)}, \label{Gsol} \\ \label{fAngel}
 f(r) &= &\ f_{0}(r) + 2 M_0 G_0 \left(\frac{G_0}{G(r)} -1 \right) \left[
1 + 
\left(\frac{G_0}{G(r)} -1 \right) \ln \left(1
-
\frac{G(r)}{G_0}\right)
\right],
\\
\Lambda(r) &=&\  
 \frac{-G(r)^2}{\ell_0^2G_0^2} \bigg[ 1 +4 \left(\frac{G_0}{G(r)} -1 \right) +
\left(5 
M_0 G_0 \frac{\ell_0^2}{r^2}
 +3\right) \left(\frac{G_0}{G(r)} -1 \right) ^2 + 
6 M_0 G_0 \frac{\ell_0^2}{r^2}
 \left(\frac{G_0}{G(r)} -1 \right)^3 
\nonumber
\\
&&\ 
+ 
2 M_0 G_0 \frac{\ell_0^2}{r^2}
\frac{G_0}{G(r)}  
\bigg(3 \left(\frac{G_0}{G(r)} -1 \right) +1\bigg) \left(\frac{G_0}{G(r)} -1 \right) ^2 \ln \left(1
-
\frac{G(r)}{G_0}\right)\bigg],
  \label{Lsol}
\eea
where $G_0,M_0,\ell_0,\epsilon$ are four integration constants. This represents a family of solutions that includes the classical BTZ black hole: the choice of the integration constants
was made by demanding that 
the classical BTZ solution is recovered when
one dimensionless constant (labeled $\epsilon$) vanishes.  
Indeed, one easily verifies that

\bea
\lim_{\epsilon\rightarrow 0}G(r)&=&G_0\hspace{.5cm},\hspace{.5cm}
\lim_{\epsilon\rightarrow 0}f(r)=-G_0 M_0+\frac{r^2}{\ell_0^2}\hspace{.5cm},\hspace{.5cm} 
\lim_{\epsilon\rightarrow 0}\Lambda(r)=-\frac{1}{\ell_0^2}
\eea
which justifies the naming of the constants ($G_0,  M_0, \Lambda_0=-1/\ell_0^2$)
in terms of their meaning in the absence of scale dependence.
%
%
The connection between  the new solution and the BTZ solution is given in terms of
the difference between the ``running'' $G(r)$ and the fixed $G_0$. One further verifies the
transition to empty $AdS_3$ space for the
classical ``mass gap'' relation $M_0 \rightarrow -\frac{1}{G_0}$
\bea
\lim_{\overset{\epsilon\rightarrow 0}{M_0 \rightarrow -1/G_0}} G(r)=G_0\hspace{.5cm}, \lim_{\overset{\epsilon\rightarrow 0}{M_0 \rightarrow -1/G_0}} f(r)= 1 + \frac{r^2}{\ell_0^2} \hspace{.5cm},\lim_{\overset{\epsilon\rightarrow 0}{M_0 \rightarrow -1/G_0}} \Lambda(r)=-\frac{1}{\ell_0^2}
\eea
Note that in this limit, all dependence on $\epsilon$ vanishes. Thus, $AdS_3$ constitutes the appropriate vacuum of the theory, which is invariant under perturbations due to the running of the couplings controlled by $\epsilon$. In terms of the RG flow, this suggests that $AdS$ would be a fixed point of the RG group.
Note further that $M_0$ is the mass of the black hole only if $\epsilon \rightarrow 0$, while for $\epsilon \neq 0$
it is much harder to determine the mass. We will come back to this point at the end of section \ref{sec_BTZsol}.

\begin{figure}[ht]
\centering
   \includegraphics[scale=1]{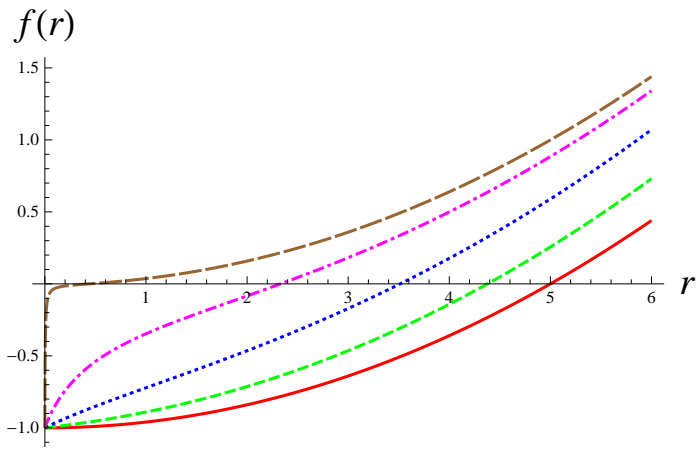}
      \caption{\label{figfr} 
    Radial dependence of the lapse function $f(r)$ for $\ell_0=5$, $G_0=1$ and $M_0=1$.
The different curves correspond to the classical case $\epsilon=0$ 
solid red line, 
$\epsilon=0.02$ 
short dashed green line,
$\epsilon=0.09$ 
dotted blue line, 
$\epsilon=0.5$ 
dot-dashed magenta line, and 
$\epsilon=100$ 
long dashed brown line.
}
\end{figure}

Since the constant $\epsilon$ controls the strength of 
the new scale dependence effects,
in some cases it is useful to treat it as small expansion parameter
\bea\label{expandeps}
G(r)&=&G_0-\epsilon \cdot (1+G_0 M_0) r+ {\mathcal{O}}(\epsilon^2),\\ \nonumber
f(r)&=&-G_0 M_0+\frac{r^2}{\ell_0^2}+2\epsilon \cdot M_0(1+G_0 M_0)r
+{\mathcal{O}}(\epsilon^2),\\ \nonumber
\Lambda(r)&=&-\frac{1}{\ell_0^2}-\epsilon \cdot \frac{2r}{\ell_0^2 G_0}(1+G_0 M_0)+ {\mathcal{O}}(\epsilon^2).
\eea
In figure \ref{figfr} the lapse function $f(r)$ is shown for different values of $\epsilon$
in comparison to the classical BTZ solution.
%

%
As can be seen in the solution \eqref{Gsol}-\eqref{Lsol}, one observes that $f(r)$ is monotonically growing for all chosen values of $\epsilon$,
that $\lim_{r\rightarrow 0}f(r)=-G_0 M_0$ independent of $\epsilon$,
and that all functions grow as $\sim r^2$ for large values of $r$.
One further notes that, even though $\lim_{\epsilon\rightarrow 0}f(r)=-G_0 M_0+r^2/\ell_0^2$, 
the first derivative of $f(r)$ at the origin is strongly dependent on $\epsilon$. As discussed below, 
the non vanishing of this derivative induced by $\epsilon$ produces a curvature singularity at the origin proportional to $\epsilon$.

\subsection{Asymptotic space-times}
\label{sec_BTZasympt}
For small radial coordinate
a new singularity appears, which is absent
in the classical BTZ solution. This can be verified by evaluating for example
the invariant Ricci scalar for the metric ansatz (\ref{lineelans})
\begin{align}\label{RicciS}
R &= - f''(r) - 2 \frac{f'(r)}{r},
\end{align}
in the limit of $r \rightarrow 0$.
One finds 
that the leading terms are
\be
R= -4M_0 \epsilon (1+G_0 M_0) \cdot  \frac{1}{r}  -
\bigg(\frac{6}{\ell_0 ^2} +10  \frac{M_0}{G_0}(1 + G_0 M_0 )^2 \epsilon ^2 \bigg) + \mathcal{O}(r^1).
\ee
This quantity is divergent for $\epsilon \neq 0$ and it is finite for $\epsilon=0$. In particular, when $\epsilon = 0$ 
one recovers the classical Ricci scalar for BTZ solution $R_{0}=-6/\ell_0^2$.

It is surprising that allowing for scale dependent couplings results in the appearance of a singularity, which was absent in the classical solution.
This effect was also observed in \cite{Contreras:2013hua,Koch:2013owa,Koch:2013rwa,Koch:2014cqa,Koch:2015nva,Gonzalez:2015upa}.
The reason is that only a very particular class of lapse functions renders the Ricci scalar finite at the origin.
This can be seen by solving the relation (\ref{RicciS}) for finite and constant $R=b$ at $r\rightarrow 0$.
The solution to this is
\be
f(r)\approx c_1+\frac{c_2}{r}-\frac{b r^2}{6}.
\ee
This shows that any lapse function which has a linear term in $r$ (or any other power $r^n$ with $n\le 2$ and $n\neq \{-1,0\}$)
necessarily produces a divergence in the Ricci scalar at $r \rightarrow 0$. 
For the solution (\ref{fAngel}), the problematic linear term can be seen in the expansion (\ref{expandeps}).

Concerning the limit $r \rightarrow \infty$,
the exact solution \eqref{fAngel} is asymptotically $AdS_3$: $f(r)\sim \frac{r^2}{\ell_0^2}$ at leading order in $r$. But although asymptotically the metric behaves as BTZ, neither $\Lambda(r)$ nor $G(r)$ mimic their BTZ analogs. Indeed, $\Lambda(r)=-3/\ell_0^2=3\Lambda_0$ at $r\rightarrow \infty$. This 'effective' cosmological constant at infinity arises from the extra term in Einstein's equation. Evaluating this term for the solution in the large $r$ regime, one has
\begin{align}\label{}
\Delta t_{\mu\nu}|_{r\rightarrow \infty}&= \frac{2}{\ell_0^2} g_{\mu\nu} .
\end{align}
When analyzing such asymptotics one has to be careful since even though $\epsilon$ is a small dimensionless parameter, other
quantities like $\epsilon r/G_0$ might actually become large at large radial coordinates. 
On the other hand, it is clear from \eqref{Gsol} that the behavior of $G(r)$ possesses two very different regimes: for $\epsilon r(1+M_0G_0)\ll G_0$, it behaves effectively as a constant $G_0$, while for $\epsilon r(1+M_0G_0)\gg G_0$ it falls to zero. 
  To consider this latter non-standard regime, one performs an expansion of the lapse function~(\ref{fAngel}), 
 where the smallness parameter is $\frac{G(r)}{G_0}\ll 1$. This yields, at first order,
\begin{align}\label{frlarge}
f|_{G\ll G_0}(r)&\approx \left( \frac{r}{\ell_0} \right)^2-\frac{2}{3}M_0G(r)\nonumber\\
&=\left( \frac{r}{\ell_0} \right)^2-\frac{2}{3}M_0G_0 \frac{G_0}{(1+M_0G_0)\epsilon r}.
\end{align}
where $\ell_0$ remains arbitrary. We shall use this approximation for the analysis of the next sections. 

\subsection{Horizon structure}
\label{sec_BTZhor}

\begin{figure}[ht]
  \includegraphics[scale=1.0]{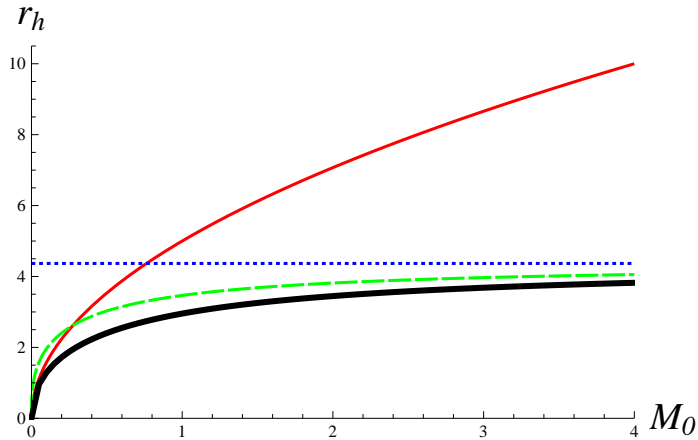}
      \caption{\label{figrHM} 
Apparent black hole horizon $r_h$ as a function of $M_0$ for $\ell_0=5$, $G_0=1$ and $\epsilon = 0.2$.
The different curves correspond to the classical case
(solid red line)
the expansion~(\ref{rh3}) for small $G/G_0$ 
(long dashed green line)
the expansion~(\ref{rh3b}) for small $G/G_0$ and large $G_0 M_0$ 
(dotted blue line),
and the numerical solution (thick solid black line).
The expansion~(\ref{horPert}) for small $\epsilon$ is not shown since for the given numerical values it would only be
reliable for very small $M_0<0.2$.
}
\end{figure}

Horizons are crucial for understanding the structure of a black hole.
Unfortunately, the zero of the lapse function~(\ref{fAngel}) implies a transcendental equation for $r$,
which can not be solved analytically. We approach this problem in three different ways:
First, we study the leading corrections with respect to the classical regime ($\epsilon$ small). 
Second, we focus on a specific region of parameter space that exhibits a particularly interesting strong scale dependence effects, namely $G(r)/G_0\ll 1$, that display some novel features.
Third, those two approaches are compared with a numerical analysis.
\begin{itemize}
\item 
Expansion in $\epsilon\ll 1$:
For weak scale dependence one can use the expansion~(\ref{expandeps}), for which
one finds the horizon 
\be\label{horPert}
r_h|_{\epsilon \ll 1}= \sqrt{G_0 M_0} \ell_0 - \epsilon \ell_0^2 M_0 (1+G_0 M_0)+ {\mathcal{O}}(\epsilon^2).
\ee
Unfortunately, an analytic result is again limited to order $\epsilon$.
One sees that the scale dependence tends to decrease the apparent horizon radius.
\item 
Expansion in $G(r_h)/G_0\ll 1$: from \eqref{Gsol}, Newton's coupling evaluated at the horizon will be much smaller than its classical value provided that
\begin{align}\label{GG1}
\epsilon r_h (1+M_0 G_0)\gg G_0
\end{align}
In this limit the horizon can be obtained from~(\ref{frlarge}).
It is the real root of
\begin{align}\label{rh3} 
r_h^3|_{G(r)/G_0\ll 1}\approx \frac{2}{3}\frac{M_0 G_0^2\ell_0^2}{(1+M_0 G_0)\epsilon}.
\end{align}
For consistency, this must satisfy \eqref{GG1}. Therefore, \eqref{GG1} and \eqref{rh3} will hold valid if the parameters satisfy the condition:
\begin{align}\label{cond}
\frac{M_0(1+M_0 G_0)^2\epsilon^2 \ell_0^2}{G_0}\gg 1
\end{align}

A particularly interesting region of the parameter space is $M_0 G_0\gg 1$. Indeed, provided that \eqref{cond} is satisfied in this limit, namely, if
\begin{align}\label{cond0}
M_0G_0\gg 1 \hspace{.5cm}\mbox{and}\hspace{.5cm} (\epsilon \ell_0)^2\gg \frac{1}{M_0^3G_0}
\end{align}

then, the radius of the horizon converges to a finite value,
\begin{align}\label{rh3b} 
r_h^3|_{G(r)/G_0\ll 1\,\& \,M_0G_0 \gg 1}\approx \frac{2}{3}\frac{G_0\ell_0^2}{\epsilon}
\end{align}

We thus see the crucial difference with its constant-coupling counterpart: for a fixed cosmological constant, the radius of the horizon remains finite as $M_0\rightarrow \infty$. 
In the light of this result one should keep in mind that $M_0$ is only the mass parameter for the classical solution and it is not the actual mass of the black hole. However, in the next section we show that for the particular case when \eqref{cond0} is valid, the physical mass indeed diverges as $M_0\rightarrow \infty$, and nevertheless the horizon remains at finite constant distance from the origin (note however that, as $\epsilon\rightarrow 0$, the 
horizon radius becomes unbounded). We shall come back to this case below. 

\item Numerical analysis:
For given values, the above analytical estimates can be compared with a numerical solution of $f(r)\overset{!}{=} 0$.
In figure \ref{figrHM} the horizon $r_h$ is shown as a function of the classical mass parameter $M_0$.
\end{itemize}
One observes that for small $M_0$, the horizon is in agreement with the classical result. 
Finally, one notes that for very large values of $M_0$ the numerical
value of the horizon saturates at constant $r_h$ which is given by the horizontal line in accordance with the $G\ll G_0$ approximation~(\ref{rh3}).
%

\subsection{Black hole thermodynamics}
\label{BTZthermo}

\begin{figure}[ht]

  \includegraphics[scale=1.0]{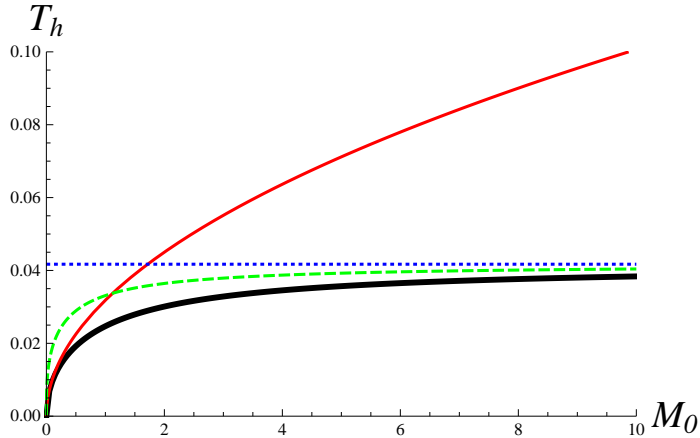}
      \caption{\label{figTH_vs_M} 
Temperature $T_h $ as a function of $M_0$ for $\ell_0=5$, $G_0=1$, and $\epsilon = 0.2$.
The different curves correspond to the classical case 
(thin solid red line),
the expansion~(\ref{Tapp}) for small $G/G_0$ 
(dashed green line),
the expansion~(\ref{Tappb}) for small $G/G_0$ and large $G_0 M_0$ 
(dotted blue line),
and the numerical solution 
(thick solid black line).
}
\end{figure}

After having gained knowledge on the horizon structure one can now turn towards
the thermodynamic properties of the solution~(\ref{Gsol}-\ref{Lsol}).
The temperature of a black hole with the metric structure~(\ref{lineelans})
is given by
\be
T=\left.\frac{1}{4 \pi}\frac{\partial f(r)}{\partial r}\right|_{r=r_h}.
\ee
Leaving the horizon radius implicit one finds
\be\label{TempGen}
T=\left.\frac{1}{2 \pi r}\frac{G_0^2 M_0}{G_0+ r \epsilon (1+G_0 M_0)}\right|_{r=r_h}.
\ee
Inserting the perturbative value for the horizon radius~(\ref{horPert})
one finds that the $ \mathcal{O}(\epsilon)$ corrections to the temperature
cancel out and that the
leading correction to the classical temperature enters at order $\epsilon^2$
\begin{align}
T |_{\epsilon\ll1}&= \frac{\sqrt{G_0 M_0}}{2 \pi \ell_0} + \mathcal{O}(\epsilon^2).
\end{align}
The transcendent structure of the solution does not allow to go straight forwardly beyond this 
$ \mathcal{O}(\epsilon)$ approximation.
However, in the opposite limit the lapse function becomes polynomial again, and one can again explore the non-classical case considered above, \eqref{cond}, namely $G\ll G_0$ and one finds from~(\ref{frlarge}) that
\begin{align}\label{Tapp}
T|_{G\ll G_0}\approx \frac{1}{4\pi} \left( 18 \frac{M_0G_0^2}{\ell_0^4(1+G_0M_0)\epsilon}  \right)^{1/3},
\end{align}
And again, within this scenario, for the particular regime of interest \eqref{cond0}, the temperature converges to a constant,
\begin{align}\label{Tappb}
T|_{G\ll G_0 \,\& \,G_0 M_0 \gg 1}\approx  \frac{1}{4\pi} \left( 18 \frac{G_0}{\ell_0^4\epsilon}  \right)^{1/3}.
\end{align}
for any finite $\epsilon,\ell_0$. 
Those analytical results can again be compared to a numerical solution of~(\ref{TempGen}).
In figure \ref{figTH_vs_M} the numerical temperature is shown as a function the parameter $M_0$
in comparison to the analytical and classical results.
One finds that the temperature behavior as a function of $M_0$ is actually a rescaled version
of the horizon radius in figure \ref{figrHM}. This is a particularity of the three dimensional case.
One further sees that for small $M_0$, the numerical curve of the new solution approaches the behavior
of the classical BTZ case. However, in the opposite limit of large $M_0$ the temperature of the new solution 
saturates at the values given by the approximations~(\ref{Tapp} and \ref{Tappb}).

Another window for the understanding the thermodynamic properties of a black hole is its entropy.
As it is well known from Brans-Dicke theory \cite{Jacobson:1993vj,Iyer:1995kg,Visser:1993nu,Creighton:1995au,Kang:1996rj}, 
the entropy of black hole solutions in $D+1$ spacetime dimensions with varying Newton's constant is given by
\begin{align}\label{}
S=\frac{1}{4}\oint_{r=r_h} d^{D-1}x\frac{\sqrt{h}}{G(x)},
\end{align}
where $h_{ij}$ is the induced metric at the horizon $r_h$. For the present spherically symmetric solution this integral is trivial. 
The induced metric for constant $t$ and $r$ slices is simply 
$ds=rd\phi$ and moreover $G(x)=G(r_h)$ is constant along the horizon due to spherical symmetry. Therefore, the entropy for this solution is
\begin{align}\label{S}
S&=\frac{A}{4G(r_h)}=\frac{A}{4G_0} \left[ 1+\frac{(1+G_0M_0)\epsilon r_h}{G_0} \right].
\end{align}
where $A=\oint_{r_h} d^2x \sqrt{h}=2\pi r_h$. We notice that the correction to the entropy is not of the form expected from many other quantum gravity programs, namely proportional to $\ln(A)$.

As expected, the entropy behaves differently in the two regimes highlighted above. 
Those two regimes can be addressed by the corresponding approximations $\epsilon \ll 1$ or $G\ll G_0$.
For very small scale dependence effects ($\epsilon \ll 1$) one finds that the $\mathcal{O}(\epsilon)$ contribution
cancels out, leaving the classical BTZ entropy up to subleading corrections
\be
S|_{\epsilon \ll 1} =
\frac{\pi}{2}\sqrt{\frac{M_0}{G_0}}\ell_0
  + \mathcal{O}(\epsilon^2).
\ee
and the entropy obeys the holographic principle according to the Bekenstein-Hawking law. 
This result can also be read directly from~(\ref{S}) in the limit of $(1+G_0 M_0)\epsilon r_h\ll G_0$. 

~

The opposite limit is however much more interesting. If the condition \eqref{cond} is satisfied, namely if the parameters satisfy \eqref{cond0}, the holographic principle is not fulfilled in its usual form. The black hole entropy is not any more proportional to the area, but rather the area times the horizon radius, which is of course due to the variation of $G(r)$.
Indeed, by inserting \eqref{GG1} into~(\ref{S}), the leading contribution to the entropy is 
\begin{align}\label{entapprox1}
S|_{G\ll G_0} &\approx \frac{A}{4G_0}\left( \frac{1+M_0G_0}{G_0} \right) \epsilon r_h \nonumber \\
&\approx
\pi  \bigg[  \frac{\ell_0 ^4 M_0 ^2 (1 + M_0 G_0)\epsilon}{18 G_0 ^2} \bigg]^{1/3}.
\end{align}
Moreover, in the regime $M_0G_0\gg 1$, this further reduces to
\begin{align}\label{entapprox2}
S|_{G\ll G_0 \,\& \,G_0 M_0 \gg 1} &\approx \frac{A}{4G_0} M_0\epsilon r_h = \pi M_0 \left(\frac{\ell_0^4 \epsilon}{18 G_0}\right)^{1/3}
\end{align}

This transition from an ``area law'' to an ``area $\times$ radius law'' is a very striking consequence of the simple
assumption of allowing for scale dependent couplings. Actually, it can be shown that this feature occurs for all spacetime dimension $d\geq 3$. Indeed, if the parameters of the scale-dependent Schwarzschild-AdS$_{d}$ solution satisfy a condition analogous to \eqref{cond}, the leading term in the entropy of the black hole scales not as $r_h^{d-1}$ but as $r_h^{d}$.  

\bigskip

\begin{figure}[ht]
  \includegraphics[scale=1]{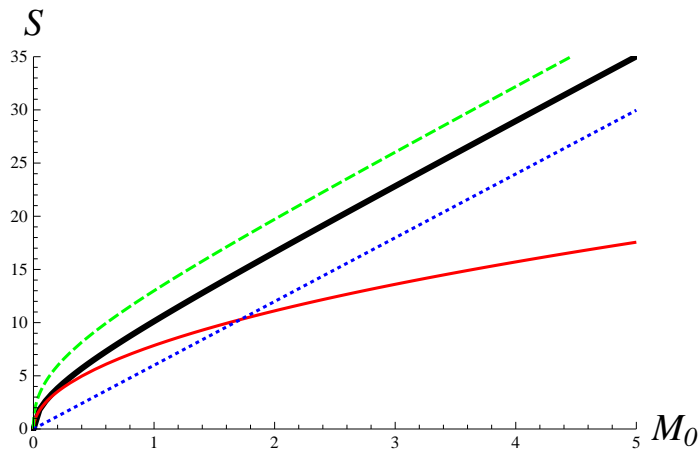}
      \caption{
 \label{figS_vs_M} 
Entropy as a function of $M_0$ for $\ell_0=5$, $G_0=1$, and $\epsilon = 0.2$. 
The different curves correspond to the classical case 
(thin solid red line),
the expansion for small $G/G_0$ 
(dashed green line),
the expansion for small $G/G_0$ and large $G_0 M_0$ 
(dotted blue line),
and the numerical solution 
(thick solid black line).
}
\end{figure}

Since the entropy~(\ref{S}) is directly given from the knowledge of the horizon radius
$r_h$ it is straight forward to implement the graphical analysis of the approximations
(\ref{rh3}, \ref{rh3b}) in comparison to the numerical result.
This is done in figure \ref{figS_vs_M}.
One notes again that the classical behavior is dominant for small $M_0$, while
for large $M_0$ a different scaling behavior appears, which is given by
the approximations \eqref{entapprox1} and \eqref{entapprox2} respectively.

Lets now come back to the physical mass of the black hole $M$.
As the discussion above showed, the classical mass parameter $M_0$ is
actually only the mass of the black hole if $G\rightarrow G_0$
\be
M|_{G\rightarrow G_0}=M_0.
\ee
In general the physical mass differs from the mass parameter, $M\neq M_0$, but it is a very difficult task to express it in a closed form. However, for the non-classical regime $G\ll G_0$ considered above, we posses analytic expressions. What is the actual mass in this regime?
This question can be answered by integrating the thermodynamic relation
\be
dM=T dS,\label{M}
\ee
which yields
\begin{align}\label{Mtrue2}
M-m=\frac{1}{4}\left( M_0+\frac{1}{G_0}-\frac{1}{3G_0} \ln(1+M_0 G_0) \right)
\end{align}
where $m$ is a constant of integration independent of $M_0$, irrelevant for these purposes. This proves the statement claimed earlier, that for fixed values of $\epsilon,\, G_0, \ell_0$ the limit of $M_0\rightarrow \infty$
implies, according to (\ref{Mtrue2}), that the physical mass grows without bound as $M \sim M_0 \rightarrow \infty$, and nevertheless the horizon converges to the finite distance given in (\ref {rh3b}).

Since our initial input was the condition (\ref{nullcond1}), 
it would be very interesting to analyse this result in future studies 
in the context of the ``Quantum Null Energy Conjecture'' \cite{Bousso:2015eda,Bousso:2015mna,Koeller:2015qmn,Bousso:2015wca,Fu:2016avb}.

\subsection{Comparison to the four dimensional solution}

In \cite{Contreras:2013hua} the scale-dependent $3+1$ black hole was considered. 
In both cases, the solution for Newton's coupling was of the form $G(r)=G_0/(1+\alpha r)$ with $\alpha$ an integration constant. This is not a coincidence: 
it can be readily shown that in any dimension the gravitational coupling takes this form \cite{prep}. 
Also, as shown above, both the three and four dimensional solutions develop a singularity of the curvature scalar at the origin. 
However, the effect mentioned after \eqref{Mtrue2} was only discussed for the three dimensional solution. 
It might actually not be the case in four dimensions, because the mass plays a very different role in the three dimensional problem as it does in higher dimensions.

\section{Summary and Conclusion}
\label{sec_sum}

A possible scale dependence of the gravitational coupling introduces an additional contribution
to the stress energy tensor of the generalized field equations (\ref{eomg}).
By imposing that the usual ``null energy condition'' is not modified by this contribution 
it is shown that in those cases any stationary solution with spherical symmetry necessarily follows the
``Schwarzschild relation'' $g_{tt}=-1/g_{rr}\equiv f(r)$.
Based on this observation
an exact spherically symmetric black hole solution
for three dimensionally gravity with scale dependent couplings is derived.
It is shown that the functional form of $f(r)$,
of Newtons coupling $G(r)$, and of the cosmological coupling  $\ell(r)$ is completely determined by the field equations.
The properties of the solution are analyzed from various perspectives.
Particular attention is dedicated to a meaningful interpretation of the integration constants
which is given in terms of the classical parameters $G_0$, $\ell_0$, $M_0$
and one additional constant $\epsilon$, that parametrizes the strength of
scale dependence.
Asymptotic spacetimes, horizon structure, and black hole thermodynamics are discussed in detail.
It is found that the large $r$ asymptotic is $AdS_3$ and that 
the $r\rightarrow 0$ asymptotic has a singular behavior.
It is found that for fixed values of $\epsilon,\, G_0, \ell_0$ the horizon radius saturates
 for $M_0 \rightarrow \infty$
to a finite value given by  (\ref{rh3b}).
Although $M_0$ is not equal to the physical mass of the black hole in general, 
in the limit of $G\ll G_0 \,\& \,G_0 M_0 \gg 1$ the physical mass $M$ grows without bound
as $M_0\rightarrow \infty$, while the radius of the horizon still converges to  (\ref{rh3b}).
The analysis of the thermodynamics showed another novel result.
Whereas for small black holes, the usual ``area law'' holds up to order ${\mathcal{O}}(\epsilon)$,
the opposite limit (which occurs when $G(r)$ deviates strongly from $G_0$)
follows an ``area $\times$ radius law''. This apparent deviation from the holographic principle
is probably the most interesting feature of this new
black hole solution with scale dependent couplings.

\section*{Acknowledgements}
We want to thank M.~Ba\~nados, C.~Armaza, and the G-HEP group at PUC
for valuable comments and discussion.
The work of B.K. was supported by the Fondecyt 1161150,
the work of I.R. was supported CONICYT-PCHA/Doctorado Nacional/2014, and
the work of A.R. was supported by
the CONICYT-PCHA/Doctorado Nacional/2015-21151658.




\bibliography{refsBTZ1}
\bibliographystyle{apsrev}

\end{document}